# An Agent-Based Approach to Component Management

**David Lillis**
University College Dublin
Belfield, Dublin 4
Ireland
+353 1 716 2908

David.Lillis@ucd.ie

**Rem Collier**
University College Dublin
Belfield, Dublin 4
Ireland
+353 1 716 2465

Rem.Collier@ucd.ie

**Mauro Dragone**
CLARITY: The Centre for Sensor Web Technologies
University College Dublin
Belfield, Dublin 4
Ireland
+353 1 716 2491
Mauro.Dragone@ucd.ie

**G.M.P. O'Hare**
CLARITY: The Centre for Sensor Web Technologies
University College Dublin
Belfield, Dublin 4
Ireland
+353 1 716 2472
Gregory.Ohare@ucd.ie

## ABSTRACT

This paper details the implementation of a software framework that aids the development of distributed and self-configurable software systems. This framework is an instance of a novel integration strategy called SoSAA (SOcially Situated Agent Architecture), which combines Component-Based Software Engineering [15] and Agent-Oriented Software Engineering, drawing its inspiration from hybrid agent control architectures. The framework defines a complete construction process by enhancing a simple component-based framework with reasoning and self-awareness capabilities through a standardized interface.

The capabilities of the resulting framework are demonstrated through its application to a non-trivial Multi Agent System (MAS). The system in question is a pre-existing Information Retrieval (IR) system that has not previously taken advantage of CBSE principles. In this paper we contrast these two systems so as to highlight the benefits of using this new hybrid approach. We also outline how component-based elements may be integrated into the Agent Factory agent-oriented application framework.

## Categories and Subject Descriptors
I.2.11 [**Artificial Intelligence**]: Distributed Artificial Intelligence – *Multiagent systems*.

## 1. Introduction
Modern distributed computing systems require powerful software frameworks to ease their development and manage their complexity. Independently from their specific nature, these systems share the need for an open and dynamic approach to system integration, as the type and the availability of their constituent parts are not stable but may change at run-time. This may occur, for example, due to changing requirements, interaction with heterogeneous/legacy systems, or network disruptions. It is thus also important to be able to change communication pathways at run-time while satisfying other run-time constrains that are dictated, for instance, by CPU and memory limitations.

The concepts that underpin component frameworka have become well-established in Component-Based Software Engineering (CBSE) [15]. Instead, researchers in this area have begun to focus on issues such as automated assembly, adaptivity, and dynamic reconfigurability, with the overarching aim of building systems that are able to meet global system requirements that may change over time [8][9]. In our mind, such aims can be addressed by adapting existing techniques from the Agent-Oriented Software Engineering (AOSE) community, such as, multi agent coordination and high-level negotiations for resource provision.

However, while AOSE has much potential for delivering open and interoperable software architectures with flexible, re-configuration capabilities, to date the take up of the approach in these domains has been limited. We argue that the limitations of AOSE are compounded by marked differences in the skill-sets and backgrounds required across the micro/functional and macro/MAS level. In particular, the emphasis of multiagent toolkits is in enabling the coordination of large scale, deliberative MASs, e.g. by deliberating the high-level goals of agents, while low-level issues arising from the interaction with the application's functionalities are often overlooked. Crucially also, the effort in standardizing the MAS level is not reflected in the way these toolkits aid integration with existing systems and infrastructures.

In order to resist this trend and work toward self-configurable distributed systems we avail of SoSAA, a software framework that integrates both CBSE and AOSE to utilise the advances already achieved in both domains. The integration of both AOSE and CBSE in a same framework permits the developing of complex MASs that needs to deal with both low level (e.g. event-based) and high level deliberative behaviors.

SoSAA incorporates modularity by applying the principles of hybrid control architectures to autonomous agents. Popularised by their use in robotics (e.g. in [6]), hybrid control architectures are layered architectures combining low-level behaviour-based systems with high-level, deliberative/procedural reasoning apparatus. From a control perspective, this enables the delegation of many of the details of the agent's control to the behaviour system, which closely monitors the agent's sensory-motor apparatus without the need to employ symbolic reasoning.

The original solution implemented in the SoSAA framework is to also apply such a hybrid integration strategy to the system's infrastructure, as illustrated by Fig. 1. SoSAA combines a low-level component-based infrastructure framework with a MAS-based high-level infrastructure framework.

The low-level framework operates by imposing clear boundaries between architectural modules (the *components*) and guiding the developers in assembling these components into a system architecture. Crucially, it then provides a computational environment to the high-level framework, which then augments its capabilities with its multi-agent organisation and goal-oriented reasoning. To this end, the SoSAA adapter provides meta-level perceptors and meta-level actuators modules, which collectively define the interface between the two layers in SoSAA.

Such an approach facilitates the integration of different functional elements in terms of independent agents, but crucially also avoids an overly rigid decomposition of the system and the overuse of symbolic interaction – two inherent risks in more traditional agent-based architectures.

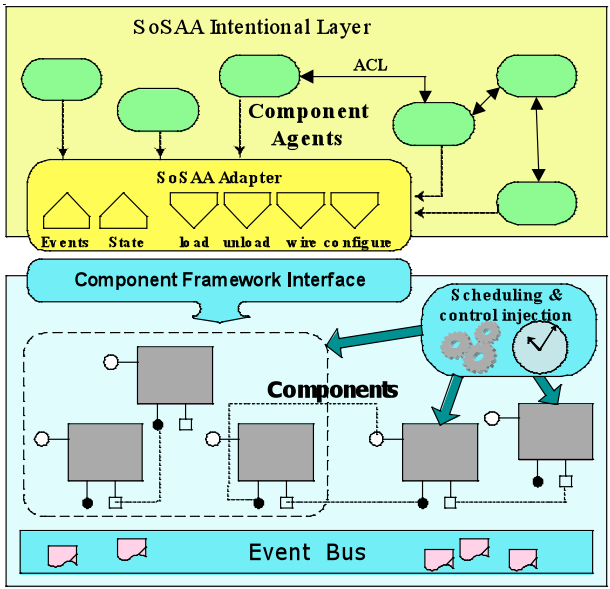

**Figure 1.** SoSAA's hybrid framework strategy.

SoSAA draws inspiration from, and contains features that are common to other systems developed by researchers in the field. Merely making use of CBSE principles to create agent systems is not a novel idea [1]. The use of "backchannels" is a key feature of RETSINA [2]. The CARTAGO system makes use of the Agents & Artifacts meta-model, utilising artifacts to provide an agent with a consistent interface through which an agent may interact with its environment [12]. We have previously performed a comparison between SoSAA and prior research such as these, which is contained in [5]. In our view, SoSAA represents a hybrid approach that combines the flexibility of component frameworks with the expressiveness and suitability of agent oriented programming languages. In particular SoSAA focuses on the exploitation of agents' goal-driven reasoning and coordination capabilities for the augmentation of different domain- and application-specific component-frameworks. This is achieved by standardizing the interaction between the agent layer and the component layer via a conceptual framework that adopts the agreed concepts that are common to component systems.

## 1.1 SoSAA Component Model

Within CBSE, domain analysis is used to capture the principal quality attributes and expresses them in form of a ***component model***. This typically provides an unambiguous description of the different component types required: their features and behavioral properties, and the set of their legitimate mutual relationships to be supported by inter-component communication channels.

In order to serve as a general-purpose infrastructure and integration framework, the SoSAA component model requires a wide range of the unifying features advanced within CBSE approaches, such as the Fractal [3] and OSGi [14] proposed standards. Collectively, these features define a generic component model that allows domain-specific specialisation while also addressing the implementation of the SoSAA hybrid framework strategy. In particular, the latter is achieved by: (i) fragmenting the fundamental mechanisms offered by the framework, and (ii) shaping their interface toward the SoSAA intentional layer as the one between a specialized class and its base class in OOP. This enables the fitting of low-level mechanisms that can be purposefully configured and overridden by the component agents in the intentional layer.

The following points present the requirements set by the SoSAA component model in greater detail:

*R1)* **Support for an extensibility set of component types**

Rather than distinguishing between rigidly pre-defined component types, SoSAA requires support for defining new component types that capture domain- and application-specific characteristics. This simplifies the development of specific applications by providing a set of primitive components that are ready to be specialized by the developer. For example, in the HOTAIR system (discussed in Section 3), components can range from active components encapsulating data-processing functionalities, to passive data components granting the access to a body of data such as task and environment-related information.

*R2)* **Recursive component context and container-type functionalities.**

The advantage of having a component context in general is the possibility to logically group components and to define context-level functions for all components in one context, beginning with their basic interface toward the component framework (e.g. *loading/unloading*, *life-cycle control*). In addition, SoSAA requires support for recursive components' contexts to hierarchically organize system components and also to provide the access to a context-level API. While a *root* context provides the main container, each component can also be a composite component by providing its own inner context to organise inter-component functionalities among its children. The basic context-level API requires container-type functionalities to load, unload, configure, and query the set of functional components loaded in the system, together with their interface requirements (in terms of *provided* and *required* collaborations).

*R3)* **Support for both *connection-* and *data-driven* component collaboration styles.**

Connection-driven interfaces are essentially procedural calls between clients and service providers. These are important to enable high performance (time-critical) quality attributes and also to allow well-defined synchronous collaborations yielding guaranteed results. As such, they can also be used for implementing behavioral coupling between components where one component uses the services exported by another component. In contrast, the data-driven composition style stresses the instantiation of indirect collaboration patterns through the transmission of either messages or events among loosely coupled components. Both mechanisms should be supported in both synchronous and asynchronous modalities, and with both unicast (one-to-one) and multicast (one-to-many) routing options. In particular, event routing needs to support prioritised dispatching of consumable events so that an event handler situated in the SoSAA intentional layer may override handlers registered within the low-level framework by: (i) registering itself as a prioritised event handler, and (ii) declaring the event consumed in order to cancel further dispatching of the event.

*R4)* **Brokering functionalities.**

These context-level functionalities act as late-binding mechanisms that can be used to defer inter-component associations by locating suitable collaboration partners for each of the collaboration styles supported by the framework. Through them, components do not need to be statically bound at design/compilation time but can be bound either at composition-time or at run-time in order to dynamically configure collaboration patterns and thus help the construction of adaptable software architectures.

*R5)* **Binding functionalities.**

Through these operations, the client-side interfaces (e.g. *service clients, event listeners, data consumers*) of one component can be programmatically bounded to server-side interfaces (e.g. *service providers, event sources, data producers*) of other components. Binding functionalities can be categorised as either explicit or implicit binding. For the former, an external controller needs to explicitly name the interfaces to be bound together, while in the latter, an internal controller will be responsible for binding a given client-side interface to any of the compatible server-side interfaces available within the context. The exact binding style used depends on the application-specific nature of the inter-component collaboration. For instance, explicit binding is required in cases where an event listener interface of one component can only receive event notifications from the event source exported by a specific component. For other applications, implicit binding is essential to provide hot-swapping (dynamic replacement of components), whereby any of the available components (e.g. exporting a stateless service) can substitute a failed one.

## 1.2 Hybrid BackChannel Management

It is relatively easy to support the aforementioned features in one single component context, as brokering and container functionalities can avail of inter-process (e.g. memory sharing) communication. To support system distribution over multiple SoSAA nodes, SoSAA advocates a hybrid communication model in the RETSINA [2] style, whereby agents can make use of *backchannels*, which allow them share information amongst themselves without the need for an Agent Communication Language (ACL). This approach assists the integration of multiple communication mechanisms (e.g. Tcp/Ip, RMI, CORBA, JMS) in one system and is thus instrumental for guaranteeing its adaptability to heterogeneous and dynamic environments.

Through its integration with a low-level component framework, SoSAA makes it easier to ground the ACL-level backchannel specifications in a set of operators that effectively manage the backchannels, as SoSAA standardises the interface opened to the agent and to its low-level components. As such, SoSAA helps to integrate legacy systems by acting as a mediator between these systems and the agent layer, especially when the low-level activities already resolve some of the issues related to the management of their backchannels. While a purely agent-based backchannel management system requires that all backchannels be managed (initiated/terminated) at the ACL level, sometimes it can be easier, advantageous, or indeed the only option, to rely upon the low-level management implemented in the component layer directly in charge of the backchannel. For instance, the successful initialization of a TCP/IP connection would already inform both participating agents when the backchannel is activated, thus rendering ACL messages such as *<connection successful>* and *<accepted client-line>* unnecessary. The SoSAA approach typically makes it easier to implement these exceptions, while also enssuring that the agent will be informed of all the backchannels, independently from their origin and history.

Figure 2 illustrates the realisation of the hybrid backchannel management in SoSAA. *Interface adapter* components provide the bridge between the standard interface classes used for inter-process component collaboration and the backchannels used to connect the corresponding components' interfaces across the network. In the example depicted in Figure 3, two components are remotely connected through a *Pull* data interface by interposing a *TcpPullServer*, which is bound to the component exporting the Pull server interface, and a *TcpPullClient* component, which is bound to the component requiring it. Figure 3 also shows how the backchannel is managed by component agents in the respective nodes. In the example, whenever the client agent wants the client component to exchange data with the server component through a TCP/IP backchannel, it needs to, respectively: (i) load the *TcpPullClient,* (ii) bind it with the client component, and (iii) configure it by passing the address of the server's node.

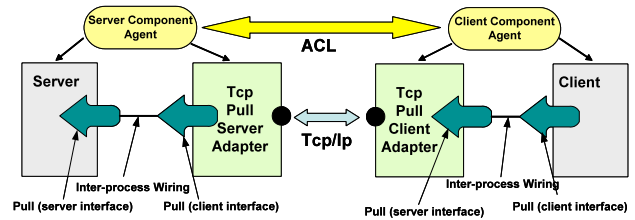

**Figure 2.** Example of backchannel management in SoSAA.

## 2. HOTAIR: Overview

To demonstrate the effectiveness and impact of using SoSAA, it is necessary to utilise it in the development of a real, non-trivial MAS. HOTAIR (Highly Organized Teams of Agents for Information Retrieval), a distributed agent-based search engine, is such a MAS, which has been built using the AFAPL2 agent programming language [13]. The system runs within the open-source Agent Factory framework, a modular and extensible framework that provides comprehensive support for the development and deployment of agent-oriented applications [4]. AFAPL2 facilitates the development of intentional agents using a variant of the widely-used BDI architecture [11]. HOTAIR had previously been developed using more traditional AOSE principles, without the use of components. It was rewritten using SoSAA so as to demonstrate the difference between the two systems. The evaluation of this is presented in Section 4.

As is common amongst Information Retrieval (IR) systems, HOTAIR consists of two distinct subsections. The Indexing Subsystem is charged with acquiring documents from a variety of sources that are stored in an index, from which retrieval can occur on the receipt of queries from users. The second subsystem, the Querying Subsystem, has the task of accepting queries from users and making use of an IR algorithm in order to return a list of documents relevant to that query.

The principal focus of this paper is on the Indexing Subsystem. The reason for this is twofold. Firstly, the two subsystems have no contact with one another, apart from the fact that they both make use of a shared index, and so they can be considered separately. Secondly, the Indexing Subsystem is not dependent on the rate at which queries arrive and so its performance can be evaluated with minimal influence from external factors.

The Indexing Subsystem consists of three processes, through which each document must travel sequentially, following a linear workflow pattern. Although full discussion of the motivations behind the choice of this specific workflow is outside the scope of this paper, the stages contained in it are outlined as follows:

The first stage of processing each document must undergo is **Data Gathering**. This is the task of identifying and downloading documents for inclusion in the index. These may be located on HTTP servers, file shares, local hard disks, FTP sites, DVDs or other sources. The type of files being downloaded is unimportant at this stage. *DataGatherer* agents are capable of downloading files and storing them on the local filesystem.

Once a document has been gathered and stored locally, it must undergo **Translation**. As documents may be in any number of file formats (e.g. HTML, PDF, Microsoft Word's .doc format), it is convenient to convert each to a common file format that is understandable by other agents within the system. This simplifies the process of Indexing, outlined below. A group of *Translator* agents are tasked with converting documents fetched by the Gatherers into a common XML-based file format that can be used to represent the contents of any such downloaded file.

*Indexing*: The final stage of processing required is to transfer the XML representation of the files to the searchable index, stored in a database. This is performed by *Indexer* agents.

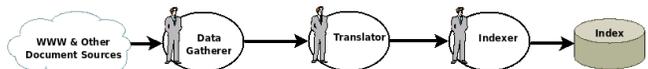

Figure 3. Indexing Subsystem Workflow

Fig. 3 displays the agents each document passes through on its way from a data source to the index. In reality, multiple agents of each type exist in the system. These form teams of agents that are responsible for performing one document processing step. They have the goal of maximizing the throughput of documents through the system, realized by ensuring that they themselves are continually engaged in finding documents (either from their original sources or other agents) and processing them.

Whenever a DataGatherer agent has located and downloaded documents, these are ready to be translated. The DataGatherer will place these documents in an output queue, from which they can be retrieved by Translators at a later stage. The agent advertises this fact by broadcasting a message to the other agents in the system to inform them of the fact that it has documents that are ready for further processing. Translators will make use of this broadcast information to make a decision about which DataGatherer it will contact in order to acquire documents that they can process. Once the translation stage has been completed, the interaction between the Translators and the Indexers operates in a similar fashion, with Translators broadcasting advertisements about documents that are ready for indexing, and Indexers contacting the relevant Translators to get these documents.

This process of Translators and Indexers automatically assigning themselves to another agent from which they can get documents operates on a greedy basis. Whenever an agent has the capacity to process more documents, it will, by default, attempt to contact the agent that had the longest output queue the last time it advertised. This tends to reduce the likelihood of individual queues growing unchecked as others are consumed.

## 2.1 Performance Management

As well as deciding for themselves where they should get documents from, the agents are also capable of following instructions sent from a *PerformanceManager* agent which may override their default behaviour. Unlike the individual processing agents, the PerformanceManager maintains an overall view of the current organisation of the system. In addition to the information broadcast about the output queues of the various agents, the PerformanceManager will also be aware of the sources each agent is using from which to fetch its documents.

This is useful for a number of reasons. By monitoring the size of the output queues of a particular team of agents, the PerformanceManager can draw conclusions about the relative performance of that team when compared with the next team in the workflow. If, for example, the output queues of the DataGatherer agents are continually growing, this is an indication that the Translators are falling behind in the rate at which they are processing documents. Such a situation is not desirable. If the DataGatherers continue fetching documents at their existing rate, their output queues will only continue to grow. This will also mean that the DataGatherers are consuming vital system resources that perhaps would be better applied to the agent teams further along the workflow, so as to ensure that documents will reach the index at a faster rate overall. Documents that have been downloaded by DataGatherers but that have not reached the index are invisible to the Querying Subsystem and as such are useless until they have undergone the Indexing process.

Similarly, Translators may be consuming documents at a faster rate than they are being downloaded by the DataGatheres. In this case, Translators continually attempting to acquire documents that are not yet available will also use system resources that would be better applied elsewhere. In situations such as these, the PerformanceManager is capable of taking a number of actions to benefit the overall performance of the system as a whole.

If the PerformanceManager believes that imbalances such as these constitute a long-term state of affairs, it has the option of requesting some agents in the over-populated group to terminate themselves. The resources freed by this action can then be used more beneficially by creating other agents as necessary. However, terminating and creating agents are computationally costly processes and so the PerformanceManager will seek to avoid this unless it is certain that such action is required. If it is unsure as to the long-term nature of a processing imbalance, it may request agents to temporarily halt in order to free up resources (such as CPU cycles) to enable other groups to catch up. In both of these situations, instructions from the PerformanceManager will override the default behaviour of the agents themselves.

In addition to balancing the performance of the agent teams so as to facilitate the steady flow of documents through the system, the PerformanceManager also has a role in balancing the system across multiple platforms. As a distributed MAS, HOTAIR facilitates the introduction of additional hardware resources. The PerformanceManager must also balance the load across machines to ensure that none is over-utilised while others are lying idle. This kind of balancing can be done by controlling the platforms on which new agents are created and also by routing documents to particular platforms (overriding the agents' default behaviour) when necessary to maintain or improve system performance.

## 2.2 Motivations for using SoSAA in HOTAIR

The existing HOTAIR system is not ideal, either from a software development or from a performance point of view.

From a software development point of view, the principal motivation behind the use of programming languages specifically designed for agent programming (such as AFAPL2) is that they aid in modeling certain features of the human cognitive process, such as beliefs, desires, intentions, roles, plans and commitments. However, if every action performed by an agent requires this type of reasoning, it tends to over-complicate the development process.

As humans, there are certain menial tasks that we can undertake without investing a significant amount of thought. Indeed we even refer to such activities as "mindless" on a regular basis. Working on a simple production line (as is the case with our agents) is one such example, and so it is desirable to separate the intentional actions from the menial tasks that are performed continually.

In the following section, we outline how we made use of the SoSAA integration strategy to perform a separation of the HOTAIR system into two layers. High-level functions such as deciding on a source from which to fetch documents, responding to and reasoning about communications received from other agents, and the management of output queues can be kept within the intentional layer. This continues to take advantage of the features of the AFAPL2 agent programming language. However, once an agent has decided to fetch documents from a particular other agent, the process of requesting, receiving and processing those documents can be performed repeatedly with minimal cognitive input. Therefore, these low-level, menial tasks are passed into the component layer, greatly simplifying the task of programming the essential intentional elements of each agent.

From a system performance point of view, inter-agent communication using Agent Communication Languages (ACLs) is a computationally expensive process [2]. In the existing HOTAIR system, all communication between agents is done using FIPA-ACL [10]. However, the introduction of components into the system allows the use of backchannels to facilitate communication between similar types of component. Additionally, components can continue their tasks of processing documents while potentially lengthy deliberation is being performed in the intentional layer.

## 3. Implementation

Microsoft's COM+ and Common Language Runtime (CLR) for the .NET platform, Sun's Java language, RMI, J2EE platform and Enterprise JavaBeans (EJB), are some of the candidate technologies to implement the SoSAA low-level component framework. However, rather than being component frameworks in their own right, these constitute component-enabling technologies that can be used to create domain specific frameworks. Also, the majority of these initiatives are biased toward business-related domains. They usually facilitate the design of multi-tier enterprise systems but provide only limited support for extension and adaptation. A notable exception in the CBSE area is the Fractal component model [3]. Specifically, Fractal introduces the notion of a component endowed with an open set of control capabilities. These are not fixed in the model but can be extended and adapted to fit the programmer's constraints and objectives.

A similar approach is adopted in the JMCF (Java Modular Component Framework), the component framework developed in conjuction with SoSAA for use across different computational environments, including resource constrained devices such as sensors and mobile phones. JMCF (illustrated in Fig. 4) is organized in a core package, which describes the framework in the form of a set of abstract interfaces, and in an implementation package. The latter includes common abstract implementation of the framework's classes as well as their domain/application- and environment--specific specialisations.

Abstract implementations of components in JMCF serve the purpose of defining *Component Type* classes that fix both the types and the implementation of components' features. The other responsibility of component type classes in JMCF is to manage the relationships with framework-type components. These are components offering framework-wide services, such as *scheduling/control-injection*, *logging* and *event-dispatching*, which can be used by the functional components defined at the application level.

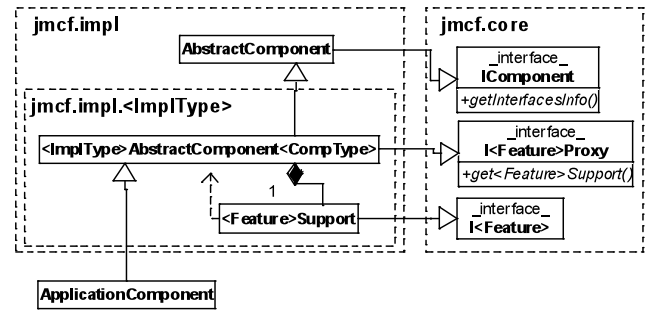

**Figure 4. JMCF** Class diagram.

Once an application's components extend a specific component type, they automatically inherit the framework mechanisms and the features supported by that component type. They are then left to declare, respectively: (i) the component's name, and (ii) the component's specific interfaces. For the latter, the component needs to override the *getInterfacesInfo* method of the *IComponent* interface. This method returns a list of *InterfaceInfo* objects, each reporting, respectively: (i) the name of the interface, (ii) the interface's collaboration style (*SERVICE*, *DATA*, *EVENT*), (iii) the interface's class (specifying either the interface implemented by the service, the type of the data, or the interface defining the source of a particular event), (iv) the interface's direction (*required/client* vs. *provided/server*), and (v) the interface's implementation (an object implementing the client or the server side of the inter-component collaboration). The component's context is responsible for the brokering and the implicit binding of components' client interfaces to suitable providers within the same context. Furthermore, in the absence of required providers within the context of the requesting component, and only if this context is not the root context, the search of a provider is extended to the context of the parent component.

JMCF comes with a package (*jmcf.impl*) of built-in component types and base-class implementations of the framework's component context class, including versions based on JavaBeans technology and a lightweight Java 1.1 compatible implementation.

## 3.1 The SoSAA Adapter

The SoSAA Adaptor acts as abridge between the low-level component framework and the higher-level AOP language. In the context of Agent Factory, this support is implemented through a combination of: a platform service, an agent module, a set of actuators and perceptors and a partial agent program that links together all the pieces and provides a basis for developing SoSAA agents. Specifically, the platform service, encapsulates the underlying component framework, and provides an interface through which that framework may be manipulated, including *loading/unloading, activation/deactivation, binding, inspection, monitoring*, and *configuration* of components.

```
ONTOLOGY sossa {
   PREDICATE activated(?cName);
   PREDICATE deactivated(?cName);
   PREDICATE focusingOn(?cName, ?type);
   PREDICATE property(?cName, ?prop, ?val);
   PREDICATE event(?cName, ?details);
   PREDICATE created(?cName);
   PREDICATE removed(?cName);
   PREDICATE component(?cName);
   PREDICATE bound(?interface1, ?interface2);
   PREDICATE clientInterface(?cName,?int,?type,?class,…);
   PREDICATE serverInterface(?cName,?int,?type,?class);
}

PERCEPTOR sosaaEventMonitor { ... }
ACTION create(?id, ?type) { ... }
ACTION remove(?id) { ... }
ACTION bind(?id1, ?iface1, ?id2, ?iface2) { ... }
ACTION configure(?id, ?param, ?value) { ... }
ACTION activate(?id) { ... }
ACTION deactivate(?id) { ... }
ACTION focus(?id) { ... }
ACTION lookup(?id) { ... }

LOAD_MODULE sosaa sosaa.module.ComponentStore;
```

**Fig 5**: Outline of the current version of the SosaaAgent.afapl2 file

Access to these operations is supported through the provision of a set of actuator units. Fig. 5 below illustrates their declaration as part of a partial AFAPL2 agent program that can be reused as a basic for creating SoSAA agents. As can be seen in this figure, this partial agent program also makes use of an agent module. Agent modules are provided by AFAPL2 to support the creation of resources that are private to a given agent. In this case, the module provides a mechanism for the agent to keep track of the components that it is interested in, and also a way of accessing the events and properties that are generated by those components. To achieve this, the *sosaaEventMonitor* perceptor has been created. This perceptor converts both events and properties into beliefs that can be used at the agent-level. It supports both basic beliefs and transformers, which generate custom beliefs.

To create a SoSAA agent, you simply import the above AFAPL2 program into your own agent program and then write your own code, using the SoSAA support where relevant. Additionally, the SoSAAService platform service must also be specified within the corresponding platform configuration file. Your AFAPL2 program must then explicitly bind to that service as is illustrated below in Fig. 6 (here it uses the service id, *af.service.sosaa*).

In this example, DataGatherer and DataQueue are Java classes that implement JMCF components that are able to process document collections and provide a queue respectively. Both classes are located within the default package (otherwise their fully qualified class name must be used). After these components are created, they are bound together so that the output interface of the DataGatherer is wired to the input interface of the DataQueue (the DataGatherer pushes new document bundles onto the queue). Finally, the agent focuses on both components, allowing it to be aware of their state and to capture any events that they raise.

The the source and the documentation of JMCF and AF ca be downloaded from http://www.agentfactory.com.

```
IMPORT sosaa.agent.SosaaAgent;
IMPORT com.agentfactory.afapl2.core.agent.BasicAgent;

COMMIT(?self, ?now, BELIEF(true),
  PAR(bindToService(af.service.sosaa),
    createHOTAIRComponent(?self, DataGatherer)
  )
);

PLAN createHOTAIRComponent(?name, ?type) {
    PRECONDITION BELIEF(true);
    POSTCONDITION BELIEF(true);

    BODY PAR(create(?name, ?type),
        specifyDataQueue(?name),
        DO_WHEN(BELIEF(dataQueueName(?qName)),
          SEQ(create(?qName, DataQueue),
            PAR(bind(?name, output, ?qName, input),
              focus(?name),
              focus(?qName)
            )
          )
        )
      );
}
```

**Fig 6**: Part of the HOTAIR application code that specifies a plan for creating a set of HOTAIR components, in this case a `DataGatherer` component and a `DataQueue` component.

## 4. Evaluation

To quantify the effects of introducing the SoSAA layer into HOTAIR, a number of experiments were conducted. The focus of these experiments was on the affect on system throughput associated with moving to a hybrid MAS/CBSE system.

## 4.1 Setup

Comparisons were made between two systems. The first is the original HOTAIR system without a component layer being utilized, as outlined in Section 3. In the following discussion, we refer to this system as "HOTAIR". The second is the updated version of HOTAIR in which the SoSAA component layer has been added to take care of low-level behaviors. This system is referred to as "HOTAIR/SoSAA".

It is important to note that moving these functions to the component layer is the only difference between the two systems. The algorithms that decide from where agents get the documents to process, and the strategies used by the PerformanceManager remain the same for both systems. Thus, these will not have any effect on the relative performance of the two systems.

Each system was seeded with two DataGatherers, each of which accessed a standard static IR dataset. The datasets used were the Cranfield corpus and the WT2G corpus from the TREC Web Track [7]. The principal motivation behind using static datasets (as opposed to gathering documents from the web, for example) was to ensure that the experimental results would not be influenced by volatile external factors such as network throughput or the response time of third party web servers.

Each system was initially run on a single machine, ceasing once 3,000 documents had been successfully indexed. This was done three times in order to reduce the influence of outliers. The results presented below use the average performance for these three runs. This process was then repeated for two, three and four machines. For these experiments with multiple machines, the Indexer agents add documents to a single searchable index.

## 4.2 Results

The results of our experiments are laid out in Fig. 7. This displays the results the experiments outlined above using both HOTAIR and HOTAIR/SoSAA. The principal result observed is that the time taken by HOTAIR/SoSAA to index 5.000 documents substantially less than for HOTAIR and that this result is consistent for all of the experiments runs. This improvement is most pronounced when the systems were run on a single platform, with HOTAIR/SoSAA using 44.48% of the time required by HOTAIR to perform the same task. Even in the case where there was the least performance improvement, namely for two platforms, the throughput of HOTAIR/SoSAA was still almost double that of HOTAIR (taking 54.15% of the time taken by HOTAIR). This result supports the assertion made in [2] that the widespread use of ACL messages to support every aspect of agent behavior is inefficient and that using backchannels to co-ordinate low-level functionality can boost a system's performance.

Despite the performance of HOTAIR/SoSAA being substantially superior to HOTAIR in all cases, it is interesting to note that in moving from one machine to two, the performance of HOTAIR/SoSAA actually declines slightly, whereas the performance of HOTAIR improves, as one would expect. The explanation for this is to be found in examining the methods of communication used by each system when running on a single host and on multiple hosts.

Running across multiple hosts introduces network overhead as an obstacle to efficient performance. With the HOTAIR system, each time a message is passed, it must be converted to an ACL and subsequently parsed on receipt. Whether local or distributed, a Message Transport Service must be present on the agent platform. The destination of the message will dictate the type of service to use, so when the system is distributed, a Local Message Transport Service is replaced with a HTTP Message Transport Service. In contrast, when HOTAIR/SoSAA is run on a single host, a component such as a Translator binding to the output queue component of a DataGatherer means that it can fetch the next documents for processing by means of a simple method invocation, which is extremely efficient. For a distributed system however, additional components are introduced at both ends of the communication: a TcpPullServer on one end and a TcpPullClient at the other. Thus, the efficiencies that are in place to aid the performance of single-host systems result in the situation where the distribution to multiple hosts causes extra overheads in addition to that of the network to be brought into the system. Despite this, it is noteworthy that the performance of HOTAIR/SoSAA is still vastly superior to that of HOTAIR and also that these extra overheads begin to be overcome with the addition of the third host, which results in performance superior to the single-host incarnation of the system.

Another notable result is an unexpected increase in processing time caused by moving from three platforms to four. It is important to note that this deterioration is observed for both systems to similar degrees (a 5% increase in processing time for HOTAIR, compared with an 8% rise for HOTAIR/SoSAA).

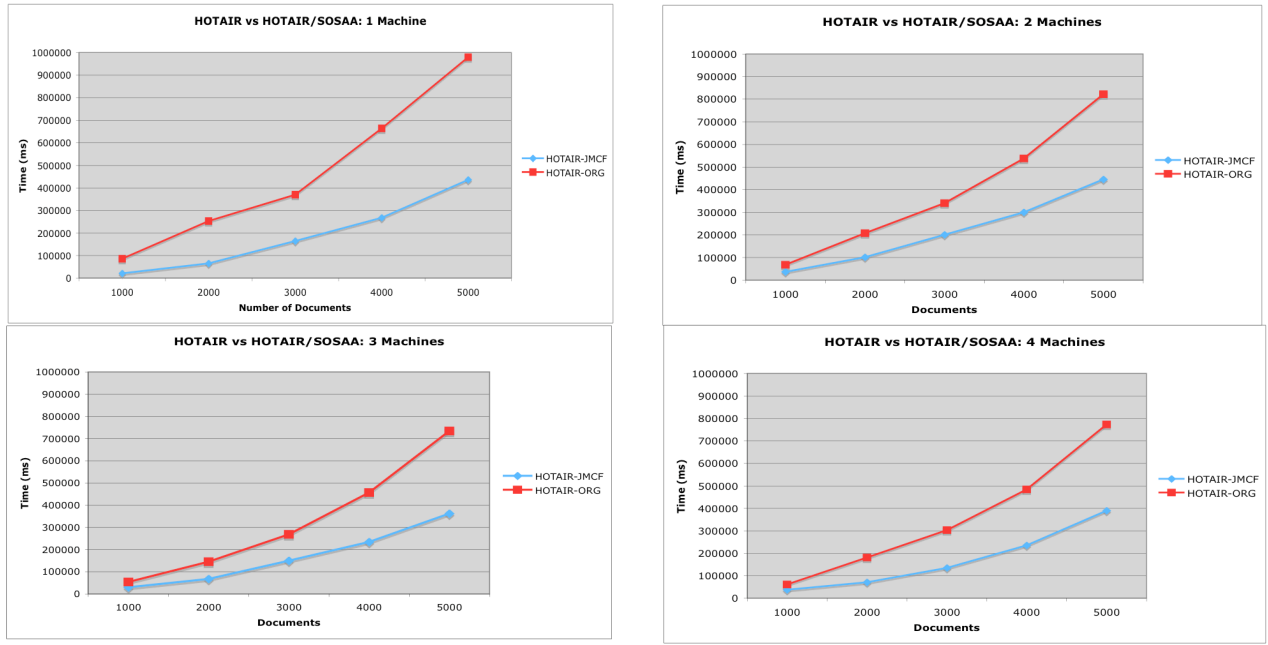

**Fig. 7**: Graphs plotting results of comparison between HOTAIR and HOTAIR/SOSAA over 1, 2, 3 and 4 machines respectively.

Because of this, we do not attribute this deterioration to shortcomings in the SoSAA implementation or integration. Rather, the PerformanceManager was observed to make decisions that did not adequately exploit the additional resources made available to it (e.g. by not creating a sufficient number of new agents to take advantage of the extra machine). Though an interesting and somewhat unwelcome result, this does not undermine the results presented. The suboptimal management agent was common to both systems and as such does not have a bearing on system performance in the comparative sense. For the experiment run with four machines, the SoSAA-enabled version of the system is still processing documents at approximately double the rate of the original HOTAIR system.

## 5. Conclusions and Future Work

In this paper, we have introduced the recently proposed SoSAA conceptual framework and its current instantiation, which combines JMCF for the component level, and Agent Factory / AFAPL2 for the agent level. Further, we have outlined the HOTAIR testbed, a distributed agent-based search engine which was developed previously using AF alone. We have used HOTAIR to evaluate our implementation of SoSAA through the creation of a SoSAA-based version of HOTAIR. In developing this new HOTAIR, we have simply replaced the underlying functional core of HOTAIR with new SoSAA components that are then managed by the re-factored AFAPL2 agents.

From a development perspective, the advantages have been:

- A clearer and cleaner separation of concerns between the underlying functionality and the agent-layer coordination mechanisms (i.e. job assignment / group management) that has improved the readability of the code base.
- The replacement of ad-hoc thread management for the underlying functionality with a more managed approach that is handled by the component framework.
- The introduction of backchannels as a mechanism for efficiently handling the transmission of job information between agents.

These benefits have led to a more efficient implementation of HOTAIR that easily outperforms the previous version, without considering any potential optimizations that could arise. Additionally, it showcases how a simple component-based framework can be enhanced through its integration with a multi agent system. This has resulted in symbiotic relationship – on the one hand, HOTAIR without components is difficult to understand and slow, and on the other hand, HOTAIR could not have been implemented with JMCF components alone because they lack reasoning and coordination capabilities. This highlights the need for frameworks such as SoSAA that provide a standardized approach to integrating CBSE and AOSE.

Future work will investigate how to inject failures to measure fault tolerance and test the hybrid backchannel management by activating different interface adapter components, e.g. replacing JMS with TCP-IP if the JMS provider seems to have failed.